\documentclass[DIV=calc,paper=a4,fontsize=11pt,twocolumn]{scrartcl} 

\usepackage[english]{babel}
\usepackage[protrusion=true,expansion=true]{microtype}
\usepackage{amsmath,amsfonts,amsthm}
\usepackage{amssymb,stmaryrd,relsize}
\usepackage[final]{graphicx}
\usepackage{xcolor}
\usepackage[normal,small,hypcap,up,labelfont=bf,textfont=it]{caption}
\usepackage{epstopdf}
\usepackage{subfig}
\usepackage{booktabs}
\usepackage{fix-cm}
\usepackage{amssymb,amsfonts}
\usepackage{dsfont}
\usepackage{bbm}
\usepackage{pstricks}
\usepackage{cite}
\usepackage[utf8]{inputenc}
\usepackage[perpage,symbol*]{footmisc}
\usepackage[varg]{txfonts}
\usepackage{balance}
\usepackage{fancyhdr}
\PassOptionsToPackage{hyphens}{url}\usepackage[pdfencoding=auto,psdextra]{hyperref}
\usepackage{bookmark}
\usepackage{verbatim}
\usepackage{fontenc}
\usepackage{cuted}

\usepackage{orcidlink}

\usepackage[shortlabels]{enumitem}

\usepackage{bm}
\usepackage{mathrsfs}

\DeclareCaptionFont{mycolor}{\color[HTML]{000000}}
\theoremstyle{definition}

\allowdisplaybreaks  

\newcommand{\U}{\uparrow}
\newcommand{\D}{\downarrow}

\usepackage{sectsty}													
\allsectionsfont{
\color[HTML]{31ADF3}\usefont{OT1}{phv}{b}{n}
}

\sectionfont{
\color[HTML]{31ADF3}\usefont{OT1}{phv}{b}{n}
}

\usepackage{fancyhdr}												
\usepackage{lettrine}
\newcommand{\initial}[1]{%
\lettrine[lines=3,lhang=0.3,nindent=0em]{
\color[HTML]{31ADF3}
{\textsf{#1}}}{}}

\usepackage{titling}															

\newcommand{\HorRule}{\color[HTML]{31ADF3}
\rule{\linewidth}{1pt}%
}

\pretitle{\vspace{-30pt} \begin{flushleft} \HorRule
\fontsize{34}{34} \usefont{OT1}{phv}{b}{n} \color[HTML]{31ADF3} \selectfont
}
\title{The Enigma of Delayed Choice Quantum Eraser}					
\posttitle{\par\end{flushleft}\vskip 0.5em}

\hypersetup{pdfborder = {0 0 0}}

\preauthor{\begin{flushleft}\large \lineskip 0.5em \usefont{OT1}{phv}{b}{sl} \color[HTML]{31ADF3}}
\author{Tabish Qureshi \orcidlink{0000-0002-8452-1078}\\[8pt]}										
\postauthor{\footnotesize \usefont{OT1}{phv}{m}{sl} \color[HTML]{000000}
Centre for Theoretical Physics, Jamia Millia Islamia, New Delhi,
India. E-mail: \href{mailto:tqureshi@jmi.ac.in}{tqureshi@jmi.ac.in}\\[10pt]		
\scriptsize\usefont{OT1}{phv}{m}{n} \color[HTML]{31ADF3}{\textbf{Editors: \emph{Stefan K. Kolev} \& \emph{Danko D. Georgiev}} }\\[5pt]
\color[HTML]{000000}{Article history: Submitted on October 19, 2025; Accepted on November 23, 2025; Published on November 23, 2025.}
\par\end{flushleft}\HorRule}

\date{}																				

\begin{document}
\maketitle
\thispagestyle{fancy} 			
\initial{T}\textbf{he delayed-choice quantum eraser represents an interesting experiment that exemplifies Bohr's principle of complementarity in a beautiful way. According to the complementarity principle, in a two-path interference experiment, the knowledge of which path was taken by the particle and the appearance of interference are mutually exclusive. Even when the which-path information is merely retained in specific quantum path-markers, without being actually read, it suffices to eliminate interference. Nevertheless, if this path information is \emph{erased} in some manner, the interference re-emerges, a phenomenon referred to as the quantum eraser. An intriguing aspect of this experiment is that if the path information is erased \emph{after} the particle has been detected on the screen, the interference still reappears, a phenomenon known as the delayed-choice quantum eraser. This observation has led to the interpretation that the particle can be influenced to exhibit characteristics of either a particle or a wave based on a decision made long after it has been registered on the screen. This idea has sparked considerable debate and discussions surrounding retrocausality. This controversy is reviewed here, and a detailed resolution provided.\\ Quanta 2025; 14: 66--74.}

\hypersetup{pdfborder = {1 1 1}}

\begin{figure}[b!]
\rule{245 pt}{0.5 pt}\\[3pt]
\raisebox{-0.2\height}{\includegraphics[width=5mm]{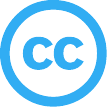}}\raisebox{-0.2\height}{\includegraphics[width=5mm]{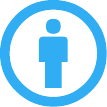}}
\footnotesize{This is an open access article distributed under the terms of the Creative Commons Attribution License \href{http://creativecommons.org/licenses/by/3.0/}{CC-BY-3.0}, which permits unrestricted use, distribution, and reproduction in any medium, provided the original author and source are credited.}
\end{figure}

\section{Introduction}

The two-slit interference experiment with quantum particles (\emph{quantons}
for short), holds a coveted position in physics. The fact the individual
massive particles, which pass through the slits one by one, accumulate
to yield an interference pattern on the screen, has intrigued researchers
since the birth of quantum mechanics. The interference characterizes the
wave nature of the quantons, whereas the knowing that a quanton passed through
a particular slit, brings out its particle nature. Niels Bohr proposed that
the two natures of quanton are mutually exclusive. In a single experiment, 
it is possible to observe only one of the two complementary aspects of the quantons.
He considered this aspect as a very fundamental feature of quantum physics,
and proposed the \emph{complementarity principle} \cite{bohr}. People believe
that the two-slit interference experiment, with the possibility of path
detection, captures the mystery of quantum mechanics in a very fundamental
way.

\begin{figure}
\centerline{\resizebox{8.0cm}{!}{\includegraphics{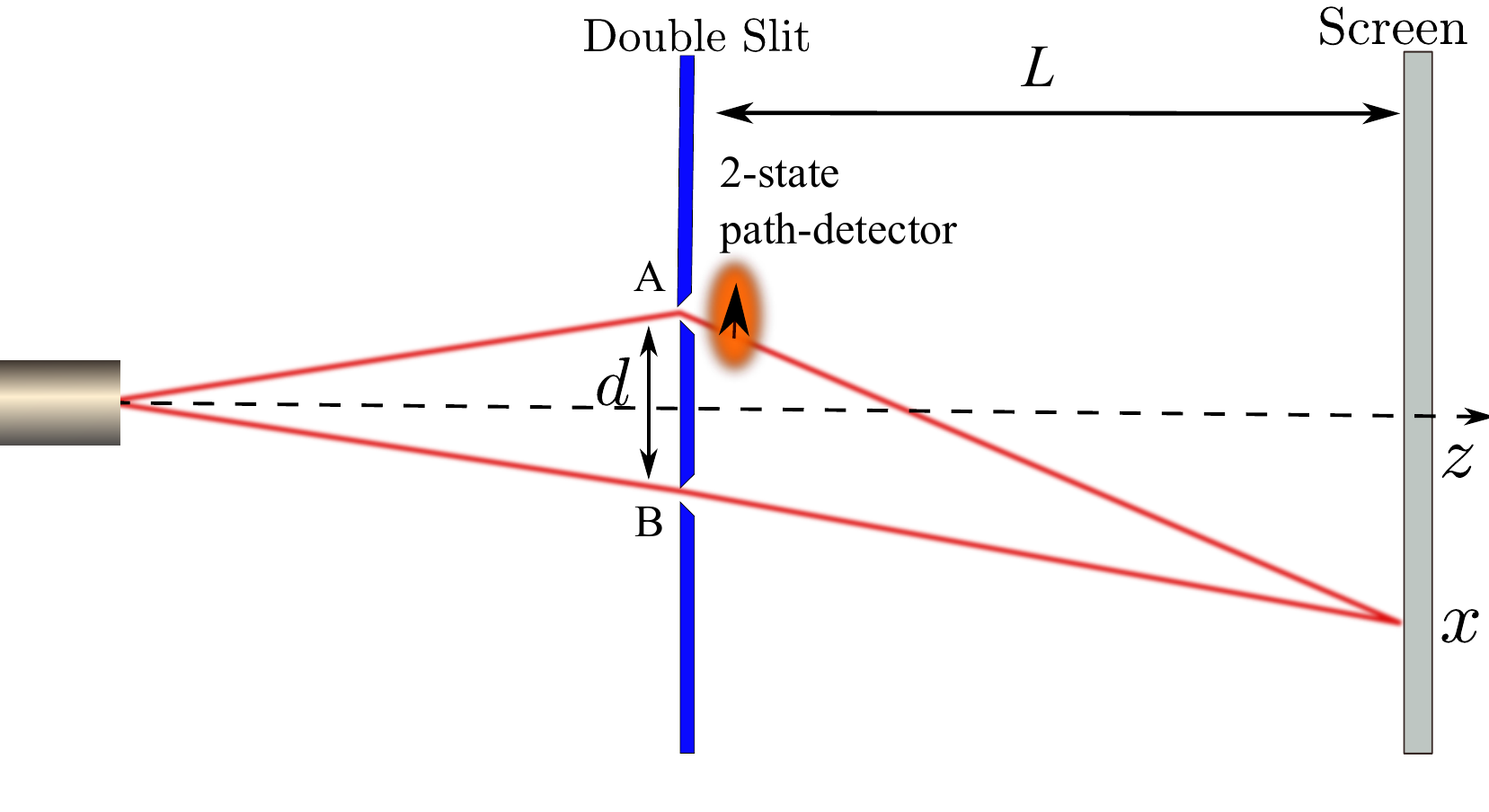}}}
\caption{Schematic diagram of a two-slit interference experiment.
There are two possible paths a quanton can take, in arriving at the screen.
}
\label{eraser2slit}
\end{figure}

Bohr's complementarity principle has stood the test of time, since its
early days when Einstein challenged it \cite{einstein}.
It has been tested in various experiments and has also been quantified
using certain duality relations \cite{englert,menon}.
Scully and Dr\"{u}hl \cite{druhl} proposed an interesting two-slit experiment
in the presence of certain quantum path marker. One could read the path
markers and force the quantons to pass through one or the other slit.
The path marker reading would indicate precisely through which slit the
quanton passed. No interference is observed for such quantons whose path
information has been extracted. Alternately one could choose to \emph{erase}
the path information
by reading out those states of the path marker which do not distinguish 
between the two paths. They showed that interference re-appears in such a
situation. In this way one could choose to make the quantons behave either as
particles, or wave. From their analysis they inferred a more dramatic result,
namely that one gets the same results even when the path marker is read
out after the quanton is detected on the screen. This appears to show that
even after the quanton has traversed the double-slit, and is detected on
the screen, one can choose to make it behave either like a wave or as
a particle. It appears that one can influence the past of the quanton.
This experiment generated a huge debate and continues to be discussed
both in the scientific and popular literature.

We will first describe the quantum eraser, and how it works, and then
discuss the various objections and interpretations. We will then explain
how the experiment should be correctly interpreted, and will show that there
is no retrocausality involved.

\section{The delayed-choice quantum eraser}

\subsection{The formulation}

Consider a two-slit interference experiment as shown in Fig.~\ref{eraser2slit}.
The quantons emerge from the source, one at a time, pass through the
double-slit, and are registered on the screen. The state of the quanton,
as it emerges from the double-slit, can be written as
\begin{equation}
|\psi_i\rangle = \tfrac{1}{\sqrt{2}}(|\psi_A\rangle + |\psi_B\rangle) ,
\end{equation}
where $|\psi_A\rangle$ ($|\psi_B\rangle$) represents the state of the quanton
if it passes through slit~A~(B). The quanton travels to the screen, and
is registered at a position $x$ on the screen, the probability density of
which is given by
\begin{eqnarray}
|\langle x|\psi_f\rangle|^2 &=& \tfrac{1}{2}\Big[|\psi_A(x)|^2 + |\psi_B(x)|^2 \nonumber\\
&&+ \psi_A(x)\psi_B^*(x) + \psi_A^*(x)\psi_B(x)\Big] ,
\end{eqnarray}
where $\psi_A(x), \psi_B(x)$  now represent the time evolved wave-functions,
$\langle x|\psi_A\rangle, \langle x|\psi_B\rangle$, of the
quanton as it travels from the double-slit to the screen. The first two
terms represent the probability density if
the quanton emerged from slit~A or slit~B. The last two terms signify the 
interference between the two amplitudes, resulting in the interference pattern 
observed in the probability distribution on the screen. This is the basic
mechanism of interference in a two-slit experiment.

\begin{figure}
\centerline{\resizebox{8.5cm}{!}{\includegraphics{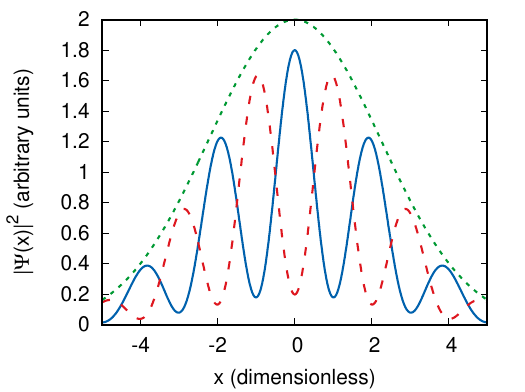}}}
\caption{A typical interference pattern in a two-slit interference in the
presence of a which-way detector. The solid line represents the
recovered interference corresponding to the path-detector state $|d_+\rangle$,
the dashed line represents the recovered interference corresponding to the
path-detector state $|d_-\rangle$. If one just detects all the quantons
without bothering about the path detector, a
washed out interference pattern (the dotted line) is obtained.
}
\label{two_patterns}
\end{figure}

Let us now introduce a path-detector at the double-slit. Without specifying
the nature of the path-detector, we just assume that it is a two-state
quantum system. If the quanton passes through slit~A~(B), the path-detector
acquires the state $|d_1\rangle$ ($|d_2\rangle$). The states
$|d_1\rangle,|d_2\rangle$ are assumed to be normalized. The combined state
of the quanton and the path-detector can now be written as
\begin{equation}
|\psi_i\rangle = \tfrac{1}{\sqrt{2}}(|\psi_A\rangle|d_1\rangle + |\psi_B\rangle|d_2\rangle) .
\label{ent1}
\end{equation}
The probability density of the quanton on the screen is now given by
\begin{eqnarray}
|\langle x|\psi_f\rangle|^2 &=& \tfrac{1}{2}\Big[|\psi_A(x)|^2 + |\psi_B(x)|^2 \nonumber\\
&&+ \psi_A(x)\psi_B^*(x)\langle d_2|d_1\rangle
+ \psi_A^*(x)\psi_B(x)\langle d_1|d_2\rangle\Big] ,\nonumber\\
\end{eqnarray}
where the interference term is now suppressed by the factor
 $|\langle d_1|d_2\rangle|$. If $|d_1\rangle,|d_2\rangle$ are orthogonal,
the interference term vanishes. By potentially measuring an observable
of the path-detector whose eigenstates are $|d_1\rangle,|d_2\rangle$,
one can unambiguously tell which slit the quanton passed through. In
accordance with the complementarity principle, one would not observe
any interference in this situation. On the other hand one could think
of another observable of the path-detector whose eigenstates are
$|d_{\pm}\rangle = \tfrac{1}{\sqrt{2}}(|d_1\rangle\pm |d_2\rangle)$.
In terms of these the entangled state (\ref{ent1}) can be written as
\begin{eqnarray}
|\psi_i\rangle &=& \tfrac{1}{2}(|\psi_A\rangle + |\psi_B\rangle)|d_+\rangle 
+ \tfrac{1}{2}(|\psi_A\rangle - |\psi_B\rangle)|d_-\rangle . \nonumber\\
\label{ent2}
\end{eqnarray}
If one looks at only those quantons for which the path-detector state is
$|d_+\rangle$, their state is given by
\begin{equation}
|\psi_+\rangle = \tfrac{1}{\sqrt{2}}(|\psi_A\rangle + |\psi_B\rangle) .
\end{equation}
This state, as we have already seen, gives rise to interference. One the
other hand, if one looks at only those quantons for which the path-detector
state is $|d_-\rangle$, their state is given by
\begin{equation}
|\psi_-\rangle = \tfrac{1}{\sqrt{2}}(|\psi_A\rangle - |\psi_B\rangle) .
\end{equation}
This state also gives rise to interference, but the interference pattern
will be shifted by a phase difference of~$\pi$, as compared to the one
corresponding to $\psi_+(x)$. The two, when added together, give no
interference. These results are summarized in Fig.~\ref{two_patterns}.
The interpretation here is simple -- once the path-detector
falls into the state $|d_+\rangle$, the potentiality of looking at 
$|d_1\rangle,|d_2\rangle$ to infer which slit the quanton went through,
is lost. In fact, finding the states $|d_{\pm}\rangle$ implies that the
\emph{quanton passed through both the slits}. This again reinforces the
complementarity principle in that the appearance of interference implies,
no which-path information can be obtained.

Now the joint probability of detecting the quanton at a position $x$ and
the path detector in a particular state, does not depend on whether one
measures the path detector before or after the quanton hitting the screen.
So it appears that one can recover interference by letting the quantons
hit the screen first, and considering them only if one gets the state
(say) $|d_-\rangle$ in a later measurement. Following the same logic,
one might infer that correlating the detected quantons with the states
$|d_2\rangle$ or $|d_2\rangle$ will tell us which slit each of them
passed through. One would not recover any interference in that situation.
This interpretation seems to imply that one could make a quanton pass 
through a single slit, or both the slits, by a delayed choice. This
broadly held view is based on the argument of Englert, Scully and Walther
\cite{esw1,esw2} which says that even in the delayed mode, the choice of
whether one wants to see $\psi_A(x),\psi_B(x)$ kind of quantons or 
$\psi_+(x),\psi_-(x)$ kind of quantons, falls to the experimenter.
The delayed-choice quantum eraser generated a huge debate which continues
to this day \cite{esw1,esw2,mohrhoff,srik,aharonov,hiley,ellerman,taming,kastnerbook1,kastnerbook2,kastner,TQ,nochoice,chiou,spekkens}.
Over the years the quantum eraser, with or
without delayed-choice, has been realized in different ways
\cite{Ma,vienna,mandel,chiao,zeilinger,kim-shih,walborn,kim,andersen,scarcelli,neves,schneider}, and several other proposals were made 
\cite{bramon,zini,barney,chianello}.
The idea of quantum eraser has also been generalized to three-path
interference \cite{3eraser}. Separated from the core issue of quantum
eraser, a novel category of delayed-choice experiments
featuring a quantum quirk has recently been investigated
\cite{terno,celeri,peruzzo,kaiser,qtwist,guo}. The objective of these
experiments was to examine the potential for a quantum superposition of
both wave and particle behaviors.

\subsection{Preliminary analysis}

It has been pointed out that the delayed-choice quantum eraser experiment
is intimately connected to the so-called Einstein--Podolsky--Rosen (EPR)
entangled state for two spin-1/2 particles \cite{esw1,kastner,TQ}.
Consider two spin-1/2 entities in a state 
\begin{equation}
|\phi\rangle = \tfrac{1}{\sqrt{2}}(|\U\rangle_1|\U\rangle_2 + |\D\rangle_1|\D\rangle_2),
\label{spinz}
\end{equation}
where labels 1 and 2 correspond to the two spins, and the states $|\U\rangle_i$ 
and $|\D\rangle_i$ represent the eigenstates of the z-component of the spins.
It is easy to see the analogy between this state and the state (\ref{ent1}).
The states corresponding to the two paths, $|\psi_A\rangle,|\psi_B\rangle$
are like eigenstates of the z-component of spin 1, and the path-detector
states $|d_1\rangle,|d_2\rangle$ are like the eigenstates of the z-component
of spin 2. Just as measuring the z-component of spin 2 can tell one about
the z-component of spin 1, looking at $|d_1\rangle,|d_2\rangle$ can tell
one which slit the quanton went through. The state (\ref{spinz}) can also
be represented as
\begin{equation}
|\phi\rangle = \tfrac{1}{\sqrt{2}}(|+\rangle_1|+\rangle_2 + |-\rangle_1|-\rangle_2),
\label{spinx}
\end{equation}
where  $|\pm\rangle_i$ represent the eigenstates of the x-component of the
spins. This state is analogous to the state (\ref{ent2}). The states
$|\pm\rangle_1$ for spin 1 are like the states $|\psi_{\pm}\rangle$ of
the quanton, which correspond to the quanton passing through both slits.
The states $|\pm\rangle_2$ for spin 2 are like the states $|d_{\pm}\rangle$
of the path-detector. Just as measuring the x-component of spin 2 can tell
one about the x-component of spin 1, looking at $|d_{\pm}\rangle$ can tell
one that the quanton went through both the slits, like a wave. The
x-component of the two spins are correlated, and so are the z-component
of the two spins. Measuring x-component of spin 2 can give no information
about the z-component of spin 1. If any third component of spin 1 is
measured, that cannot give any information about x- or z-component of
spin 2. Measuring any third component of spin 1 first, and then choosing
to measure x- or z-component of spin 2 subsequently, to find out what
the x- or z-component of spin 1 was, is simply nonsensical. Analogously
in the delayed-choice quantum eraser, if the quanton has already hit the
screen, the quanton has been measured in some other basis. Measuring 
$|d_{1,2}\rangle$ or $|d_{\pm}\rangle$ subsequently to infer whether the
quanton went through a particular slit, or both slits, is also nonsensical.
This is the fundamental mistake in the prevalent interpretation of the 
delayed-choice quantum eraser \cite{esw1,esw2}.

\section{A $n-$channel quantum eraser}

It has been demonstrated earlier that the delayed-choice quantum eraser
can be understood better if one uses a Mach--Zehnder interferometer \cite{MZ-1,MZ-2} instead
of a two-slit setup \cite{TQ,nochoice}. In a Mach--Zehnder interferometer 
there are two output detectors. The default interference is represented
by one particular detector detecting all the quantons, and the other one
detecting none. The complementary interference is represented by the
other detector registering all the quantons, and the first one detecting
none. So path-detector state $|d_+\rangle$ will correspond to all quantons
going to the first detector, and $|d_-\rangle$ will correspond to all of
them going to the second detector. Here it is obvious that, in the delayed
mode, each detector clicking will correspond to either $|d_+\rangle$ or
$|d_-\rangle$ state of the path-detector, which means that the quanton
followed both the paths, and not just one of the two \cite{TQ,nochoice}.

\begin{figure}
\centerline{\resizebox{8.0cm}{!}{\includegraphics{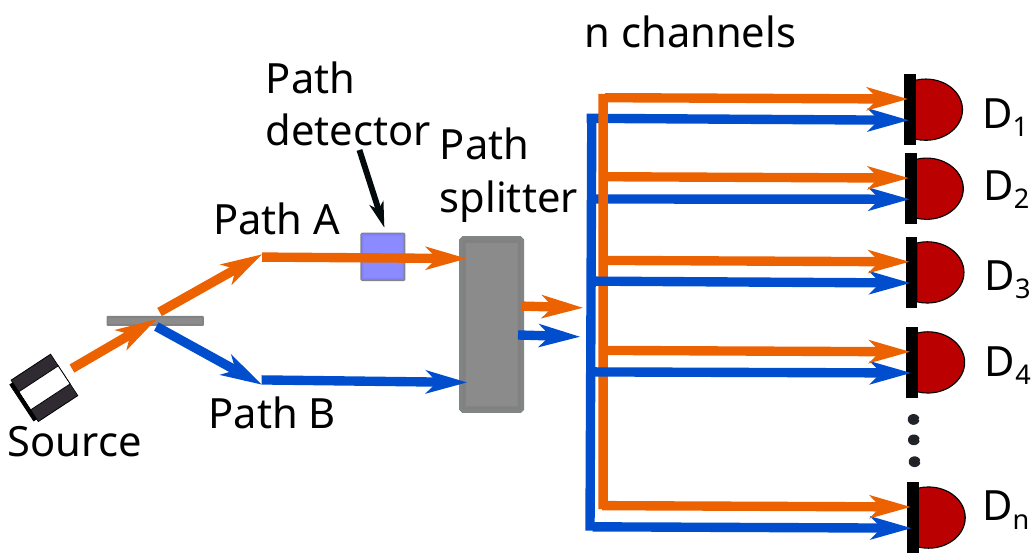}}}
\caption{Schematic diagram of a two-path, $n-$channel interference experiment.
There are two possible paths a quanton can take, in arriving at the $n$
output detectors.
}
\label{neraser}
\end{figure}

\begin{figure}
\centerline{\resizebox{8.0cm}{!}{\includegraphics{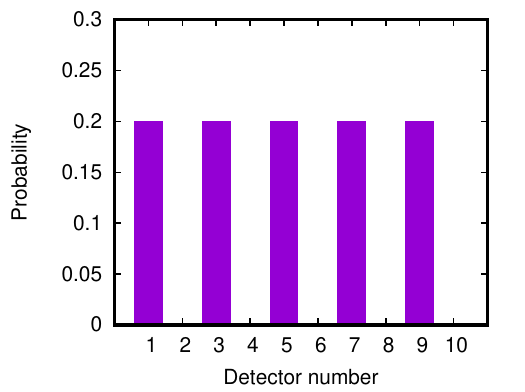}}}
\caption{A typical interference pattern for $n=10$ channels. All the quantons
land only at odd numbered detectors, and none at even numbered ones. This
represents a fringe pattern.
}
\label{ninterf}
\end{figure}

In order to make a closer correspondence to the two-slit experiment,
here we propose a discrete version of quantum eraser in a more general
setting. Consider that the quanton can take two possible paths,
and each path is then split into $n$ common channels (see Fig.~\ref{neraser}).
So a quanton taking
path A is equally likely to go to any of the $n$ output channels. The
same holds for a quanton taking path B. It is easy to imagine that 
$n=2$ corresponds to the Mach--Zehnder interferometer, where each path is
split into two output channels. Let us assume that a quanton emitted
from the source is split into a superposition of two paths, A and B.
The state of the quanton may be written as
\begin{equation}
|\psi_i\rangle = \tfrac{1}{\sqrt{2}}(|\psi_A\rangle + |\psi_B\rangle) .
\label{entn0}
\end{equation}
Quanton in each path encounters a path-splitter which splits it into
a superposition of $n$ channels, each ending in a detector.
The action of the path-splitter on the two states can be captured by
the effect of a unitary operator $\mathbf{U_{PS}}$ in the following way,
\begin{eqnarray}
\mathbf{U_{PS}}|\psi_A\rangle &=& \tfrac{1}{\sqrt{n}}\sum_{j=1}^n
e^{i\theta_j}|D_j\rangle\nonumber\\
\mathbf{U_{PS}}|\psi_B\rangle &=& \tfrac{1}{\sqrt{n}}\sum_{k=1}^n
e^{i\phi_k}|D_k\rangle,
\label{nsplit}
\end{eqnarray}
where $\theta_m, \phi_m$ are the phases picked up by the quanton
in arriving at the detector $D_m$, from paths A and B, respectively.
The state of a quanton, when it goes through the m'th channel, and
arrives at the detector $D_m$, is represented by $|D_m\rangle$.
Thus the state of a quanton passing through the two paths A and B,
and arriving at the final detectors, is given by
\begin{eqnarray}
|\psi_f\rangle &=& \mathbf{U_{PS}}\tfrac{1}{\sqrt{2}}(|\psi_A\rangle + |\psi_B\rangle)\nonumber\\
&=& \tfrac{1}{\sqrt{2n}}\sum_{j=1}^n
(e^{i\theta_j}+e^{i\phi_j})|D_j\rangle .
\label{nstate}
\end{eqnarray}
In a simplest case we assume that $\theta_j=0$ for all $j$, and
$\phi_j=0$ for odd $j$'s, and $\phi_j=\pi$ for even $j$'s. From
(\ref{nstate}) one can see that the amplitude for $|D_j\rangle$
will be $\sqrt{2/n}$ for odd $j$'s, and zero for even $j$'s.
Consequently all quantons will go to odd numbered detectors, and none to
even numbered ones. These correspond to the bright and dark fringes
in a two-slit interference experiment (see Fig.~\ref{ninterf}).

\begin{figure}
\centerline{\resizebox{8.0cm}{!}{\includegraphics{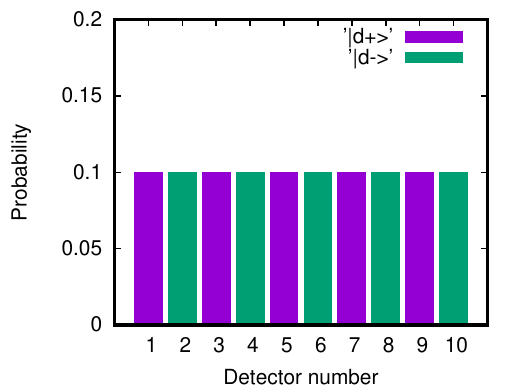}}}
\caption{Recovered interference patterns for $n=10$ channels, in the presence
of a path detector. Corresponding to the path-detector state $|d_+\rangle$
all the quantons land only at odd numbered detectors. Corresponding to the
path-detector state $|d_-\rangle$ all the quantons land only at even numbered
detectors. 
}
\label{ninterfe}
\end{figure}

Now in the presence of a path-detector in the path of the quanton, the initial
state is
\begin{equation}
|\psi_i\rangle = \tfrac{1}{\sqrt{2}}(|\psi_A\rangle|d_1\rangle + |\psi_B\rangle|d_2\rangle) .
\label{entn}
\end{equation}
The final state is then given by
\begin{eqnarray}
|\psi_f\rangle &=& \mathbf{U_{PS}}\tfrac{1}{\sqrt{2}}(|\psi_A\rangle|d_1\rangle + |\psi_B\rangle|d_2\rangle)\nonumber\\
&=& \tfrac{1}{\sqrt{2n}}\sum_{j=1}^n
(e^{i\theta_j}|d_1\rangle+e^{i\phi_j}|d_2\rangle)|D_j\rangle .
\label{nstate1}
\end{eqnarray}
With the phases as before, $\theta_j=0$ for all $j$, and
$\phi_j=0$ for odd $j$'s, and $\phi_j=\pi$ for even $j$'s, we
find that the probablity of all output detectors registering the quanton
is the same,
$\tfrac{1}{2n}\big|e^{i\theta_j}|d_1\rangle+e^{i\phi_j}|d_2\rangle\big|^2=\tfrac{1}{n}$.
This implies no interference. Now the state (\ref{nstate1}) can
also be written as
\begin{eqnarray}
|\psi_f\rangle &=& \tfrac{1}{2\sqrt{n}}\sum_{j=1}^n
(e^{i\theta_j}+e^{i\phi_j})|d_+\rangle)|D_j\rangle \nonumber\\
&&+ \tfrac{1}{2\sqrt{n}}\sum_{j=1}^n 
(e^{i\theta_j}-e^{i\phi_j})|d_-\rangle)|D_j\rangle .
\label{nstate2}
\end{eqnarray}
The quantons, for which the path-detector state is found to be $|d_+\rangle$, 
will be in the state
\begin{eqnarray}
\langle d_+|\psi_f\rangle &=& \tfrac{1}{2\sqrt{n}}\sum_{j=1}^n
(e^{i\theta_j}+e^{i\phi_j})\langle d_+|d_+\rangle)|D_j\rangle \nonumber\\
&&+ \tfrac{1}{2\sqrt{n}}\sum_{j=1}^n 
(e^{i\theta_j}-e^{i\phi_j})\langle d_+|d_-\rangle)|D_j\rangle \nonumber\\
&=& \tfrac{1}{2\sqrt{n}}\sum_{j=1}^n
(e^{i\theta_j}+e^{i\phi_j})|D_j\rangle 
\label{nstate+}
\end{eqnarray}
Similarly the quantons, for which the path-detector state is found to be
$|d_-\rangle$, will be in the state
\begin{eqnarray}
\langle d_-|\psi_f\rangle &=& \tfrac{1}{2\sqrt{n}}\sum_{j=1}^n
(e^{i\theta_j}-e^{i\phi_j})|D_j\rangle  .
\label{nstate-}
\end{eqnarray}
Considering that $\theta_j=0$ for all $j$, and
$\phi_j=0$ for odd $j$'s, and $\phi_j=\pi$ for even $j$'s,
it is straightforward to see that the quantons in the state (\ref{nstate+})
will all land on odd numbered detectors, and the ones in the state
(\ref{nstate-}) will all land on even numbered detectors. The outcome is
illustrated in Fig.~\ref{ninterfe}. It can be interpreted as a discrete
version of Fig.~\ref{two_patterns}.

What is most interesting in this $n-$channel quantum eraser is that in the
\emph{delayed mode}, every quanton detected at an odd numbered detector will
throw the path-detector in a \emph{definite} state $|d_+\rangle$, and every
quanton detected at an even numbered detector will throw it 
in a \emph{definite} state $|d_-\rangle$. And both these path-detector
states correspond to the quanton following both the paths, hence behaving like
a wave. So in the delayed mode there is no choice for the experimenter.
Every registered quanton follows both the paths, and the which-way information
is erased. This contradicts the widely held belief \cite{esw1,esw2}.

\section{The two-slit quantum eraser}

The lingering question is whether we can derive the same conclusion for a
two-slit quantum eraser as we did for the $n-$channel quantum eraser. We will
address this question in the present section. The first thing to recognize
is that in a two-slit quantum eraser the two complementary interference
patters, depicted in Fig.~\ref{two_patterns}, are not the only ones that
can be recovered, and $|d_{\pm}\rangle$ are not the only two path-detector
states that can be used for quantum erasing. In fact there exist an 
infinite number of mutually unbiased basis sets for the path detectors,
$|d^{\theta}_{\pm}\rangle = \frac{1}{\sqrt{2}}(e^{i\theta}|d_1\rangle \pm
e^{-i\theta}|d_2\rangle)$, 
which can be used for quantum erasing, where $\theta$ is an arbitrary phase
factor. The recovered interference pattern corresponding to
$|d^{\theta}_{+}\rangle$ will be shifted as compared to the recovered
pattern corresponding to $|d_{+}\rangle$. However in the delayed mode, as
the quanton is registered on the screen first, there is no reason why
the states $|d_{\pm}\rangle$ should emerge on their own.

In order to understand what happens in a two-slit quantum eraser, we
approach it in a manner akin to the $n-$channel quantum eraser examined
in the previous discussion. Nevertheless, in this case, the channels
(as well as the final detectors) are neither discrete nor finite.
Rather the screen represents an infinite number of
positions $x$ at which the quanton can arrive. Also, in this case there
is no path-splitter that is employed. It is the Schr\"odinger evolution of
the initially localized wave-packet emerging from a slit, which makes it
spread over an infinite set of positions at the screen. The
path-splitter action depicted by (\ref{nsplit}) is then modified to
a continuous scenario as
\begin{eqnarray}
\mathbf{U_{PS}}|\psi_A\rangle &=& \int \psi(x) e^{i\theta_x}|x\rangle dx
\nonumber\\
\mathbf{U_{PS}}|\psi_B\rangle &=& \int \psi(x) e^{i\phi_x}|x\rangle dx,
\label{xsplit}
\end{eqnarray}
where $|x\rangle$ represents a position eigenstate, and $\psi(x)$ is an
envelope function which is approximately assumed to be the same for the
two states. In the presence of a path-detector, the final state at the
screen is given by
\begin{eqnarray}
|\psi_f\rangle &=& \mathbf{U_{PS}}\tfrac{1}{\sqrt{2}}(|\psi_A\rangle|d_1\rangle
+ |\psi_B\rangle|d_2\rangle) \nonumber\\
&=& \tfrac{1}{\sqrt{2}}\int \psi(x)(e^{i\theta_x}|d_1\rangle 
+ e^{i\phi_x}|d_2\rangle)|x\rangle dx. 
\end{eqnarray}
This represents no interference because $|\langle x|\psi_f\rangle|^2 = 
|\psi(x)|^2$, due to the orthogonality of $|d_1\rangle,|d_2\rangle$.
In a two-slit experiment, as shown in Fig.~\ref{eraser2slit}, the phases
are known to be $\theta_x = \pi xd/\lambda L$ and $\phi_x = -\theta_x
= -\pi xd/\lambda L$, where $\lambda$ is the wavelength associated with the
quanton. The above state can then be written as
\begin{eqnarray}
|\psi_f\rangle &=& \tfrac{1}{\sqrt{2}}\int \psi(x)(e^{i\theta_x}|d_1\rangle 
+ e^{-i\theta_x}|d_2\rangle)|x\rangle dx. 
\end{eqnarray}
Now if we choose a basis, for the path-detector states, given by
$|d^\theta_{\pm}\rangle = \tfrac{1}{\sqrt{2}}(e^{i\theta}|d_1\rangle
+ e^{-i\theta}|d_2\rangle$, the above state can be written as
\begin{eqnarray}
|\psi_f\rangle &=& \tfrac{1}{2}\int \psi(x)(e^{i(\theta_x-\theta)} 
+ e^{-i(\theta_x-\theta)})|d^{\theta}_+\rangle|x\rangle dx \nonumber\\
&& + \tfrac{1}{2}\int \psi(x)(e^{i(\theta_x-\theta)} 
- e^{-i(\theta_x-\theta)})|d^{\theta}_-\rangle|x\rangle dx \nonumber\\
&=& \int \psi(x)\cos(\theta_x-\theta)|d^{\theta}_+\rangle|x\rangle dx \nonumber\\
&& + \int \psi(x)i\sin(\theta_x-\theta)|d^{\theta}_-\rangle|x\rangle dx .
\label{xinterf}
\end{eqnarray}
Path information can be erased by choosing the path-detector state
$|d^{\theta}_+\rangle$, which yields the probability density of finding
the quanton at a position $x$
\begin{eqnarray}
|\langle x|\otimes\langle d^{\theta}_+|\psi_f\rangle|^2 &=& \tfrac{1}{2}|\psi(x)|^2
[1 + \cos(\tfrac{2\pi xd}{\lambda L}-2\theta)]~~~~~~
\label{pattern+}
\end{eqnarray}
The probability density of quantons coincident with $|d^{\theta}_-\rangle$ is
given by
\begin{eqnarray}
|\langle x|\otimes\langle d^{\theta}_-|\psi_f\rangle|^2 &=& \tfrac{1}{2}|\psi(x)|^2
[1 - \cos(\tfrac{2\pi xd}{\lambda L}-2\theta)] .~~~~~~
\label{pattern-}
\end{eqnarray}
Eqns. (\ref{pattern+}) and (\ref{pattern-}) represent two complementary
interference patterns similar to those depicted in Fig.~\ref{two_patterns},
but a little shifted if $\theta$ is nonzero.

Now we are in a position to understand what happens in the delayed mode.
From (\ref{xinterf}) one can see that when a quanton lands at a position
$x$, both $|d^{\theta}_+\rangle$ and $|d^{\theta}_-\rangle$ have nonzero
probabilities of occurring. However, if the path-detector basis is chosen
such that $\theta=\theta_x$, for that particular position $x$, the combined
state (\ref{xinterf}) becomes
\begin{equation}
|\psi_f\rangle = \int \psi(x) |d^{\theta_x}_+\rangle|x\rangle dx ,
\end{equation}
which means that the path-detector comes to a \emph{definite} state
$|d^{\theta_x}_+\rangle$. This in turns means that the path information
is erased, and the quanton followed both paths. For each position $x$ at
which a quanton arrives, there is a corresponding basis in which the
path detector will be in a definite state. This indicates that even
in the two-slit delayed-choice quantum eraser, the
quanton always traverses both paths, and the which-path information
is always erased. There is no choice
left for the experimenter to recover the path information.

\section{Discussion}

We first analyzed a thought implementation of quantum eraser in a
two-path interference experiment where each path is split into $n$ channels.
The two amplitudes from the two paths interfere constructively in some
channels and destructively in others. This may be interpreted as a discrete
version of the conventional two-slit interference experiment. For $n=2$
it reduces to the Mach--Zehnder interferometer. A path-detector is added to
the setup to obtain information on which of the two paths a quanton followed.
It is demonstrated that in the delayed mode a quanton landing at an odd
(even) numbered detector leaves the path-detector in the state
$|d_+\rangle$ ($|d_-\rangle$). Both these states correspond to the quanton
following both paths, like a wave. This conclusively shows that in the
delayed mode of the quantum eraser, the quanton always follows both the
paths, and the which-path information gets erased. 

Next we analyzed the conventional two-slit delayed-choice quantum eraser
using the same methodology that was used to study the $n-$channel quantum
eraser. It is a bit more involved than the $n-$channel quantum eraser
because the phase picked up by the quanton in arriving at a position $x$
on the screen is different for each position. Nevertheless, in the delayed
mode, every quanton still passes through both the slits, and this information
is registered in a \emph{definite} state of the path detector. This path
detector state depends on the position where the quanton lands, and is
given by $\tfrac{1}{\sqrt{2}}(e^{i\theta}|d_1\rangle + e^{-i\theta}|d_2\rangle$,
where $\theta = \pi x/w$, $w$ being the fringe-width of the two-slit
interference $w=\lambda L/d$. This can be verified in an experiment by
first locating the precise position of the quanton, inferring the value
of $\theta$ from that, and then probing the
path detector in the basis formed by the states
$\tfrac{1}{\sqrt{2}}(e^{i\theta}|d_1\rangle \pm e^{-i\theta}|d_2\rangle$.
Previous authors \cite{esw1,kastner} correctly identified the analogy
between the quantum eraser and the an EPR pair of \mbox{spin-1/2}, but missed the
point that one needs to consider other mutually unbiased basis sets of the
path detector to correctly interpret the experiment.

\pagebreak
Finally we would like to mention some important interpretations of
the delayed-choice quantum eraser, that our analysis provides.
\begin{itemize}
\item If one measures the which-way detector first,
one may choose to obtain which-way information, or erase the which-way
information by looking at two different sets of basis states.
\item If the which-way detector is not measured,
the quanton follows both the paths, always.
\item In the delayed mode, the which-way information is erased for every
quanton, yet the interference is lost. The absence of interference in this
context does not imply that any which-way information exists.
\item In the delayed mode, for every detected quanton the
path detector is left in a definite state. This state informs us not only
that the quanton traversed both paths, but it also provides precise
information regarding the phase difference between the two paths.
\item Nonetheless, the interference pattern, of any specific form selected,
can be retrieved in the delayed mode by opting to measure the path detector
in the basis pertinent to the selection, and by correlating the registered
quantons with various detector states. However, these recovered patterns 
should not be interpreted as telling us how the quanton traveled the
two paths. That information is there in the definite state that the
path detector is left in.
\end{itemize}

\balance
\interlinepenalty=10000

\end{document}